\documentclass{iau}
 \usepackage{graphicx} 

\title[IAUS291.~~Annular gap model for pulsars] 
{Annular gap model for multi-wavelength pulsed emission from
  young and millisecond pulsars} 

\author[Y. J. Du \& G. J. Qiao]  
{YuanJie Du$^1$
 \and GuoJun Qiao$^2$}

\affiliation{$^1$National Space Science Center, Chinese Academy of
  Sciences, \\NO.1 Nanertiao, Zhongguancun, Haidian district, Beijing
  100190, China \\ email: {\tt duyj@nssc.ac.cn} \\[\affilskip]
  $^2$School of Physics, Peking University, Beijing 100871, China
  \\email: {\tt gjn@pku.edu.cn}}

\pubyear{2012}
\volume{291}  
\jname{\mbox{Neutron Stars and Pulsars: Challenges and Opportunities
    after 80 years}} 
\editors{J. van Leeuwen, ed.}

\begin{document}

\maketitle

\begin{abstract}
The multi-wavelength pulsed emission from young pulsars and
millisecond pulsars can be well modeled with the single-pole
3-dimension annular gap and core gap model. To distinguish our single
magnetic pole model from two-pole models (e.g.\ outer gap model and
two-pole caustic model), the convincing values of the magnetic
inclination angle and the viewing angle will play a key role. 
\end{abstract}


\firstsection 

\section{Introduction}

Pulsars are fascinating astronomical objects in the universe. Many
pulsars, including young normal pulsars and millisecond pulsars
(MSPs), radiate multi-wavelength pulsed emission, which have not been
completely understood.

High energy emission (e.g. $\gamma$-ray emission) from pulsars takes
away a significant fraction of the rotational energy. Thanks to the
launching of \textit{Astro-rivelatore Gamma a Immagini LEggero} (AGILE) and
the \textit{Fermi Gamma-ray Space Telescope} (FGST), more than one hundred new
$\gamma$-ray pulsars have been discovered in the last year, including
gamma-ray only pulsars and a new population of millisecond pulsars
(\cite[Abdo et al.\ 2009, Abdo et al.\ 2010a, Pellizzoni et al.\ 2009]{msp-sci,catalog,pel09}).
From observations, MSPs are analogous to young pulsars, which have
multi-wavelength pulsed emission from radio ($10^{-6}$\,eV) to
$\gamma$-ray band. Do MSPs and young pulsars share a simple model that
contains simlar emission region and acceleration mechanism to
self-consistently explain their multi-wavelength emission?

\begin{figure*}
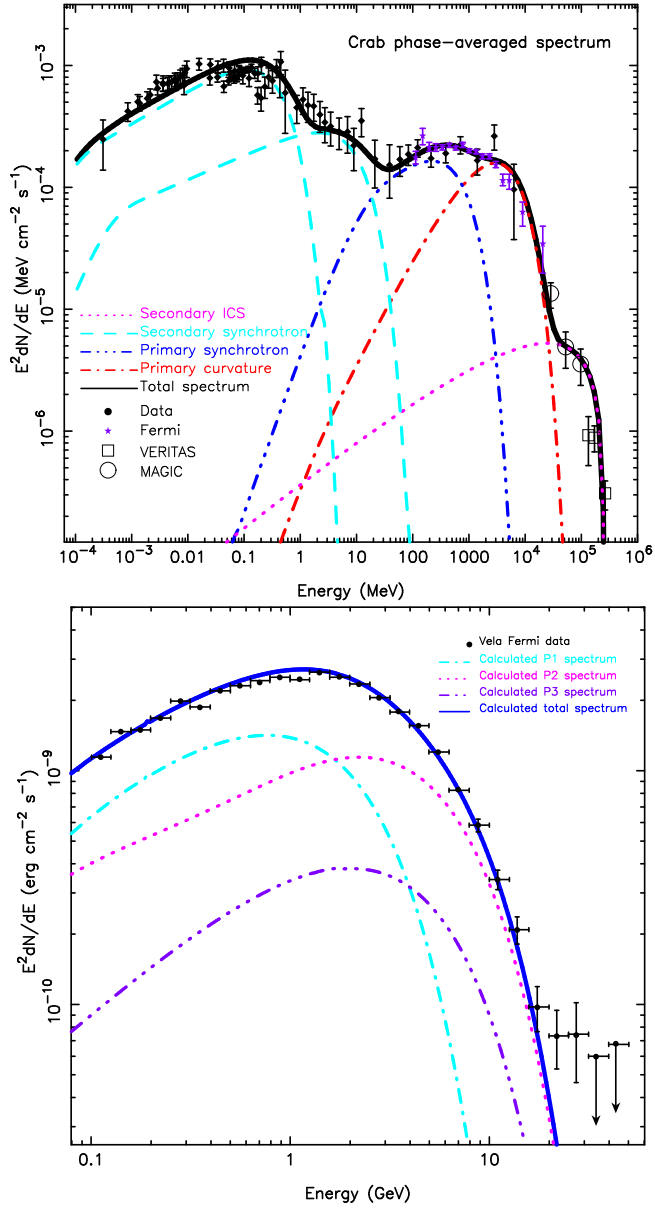

\begin{center}
\includegraphics[angle=0,scale=.57]{crab_PAS_MV0.eps}
\includegraphics[angle=0,scale=.57]{vela_Ga_PAS.eps}
\caption{Upper panel: the modeled radio-to-TeV phase-averaged spectrum
  for the Crab pulsar (adopted from \cite[Du et
    al.\ 2012]{crab}). Bottom panel: the modeled $\gamma$-ray
  phase-averaged spectrum for the Vela pulsar (adopted from \cite[Du
    et al.\ 2011]{vela}).} \label{PAS}
\end{center}
\end{figure*}
%
Some models have been proposed to explain pulsar's multi-wavelength
emission, as we will summarize below, differing on the acceleration
region of the primary particles and the mechanism for the production
of the high energy photons.
Initially aiming to explain the high-energy pulsed emission from young
pulsars, four traditional physical or geometrical magnetospheric
models which have been proposed to explain pulsed $\gamma$-ray
emission of pulsars: the polar cap model (\cite[Daugherty \& Harding
  1994]{1994ApJ...429..325D}), the outer gap model (\cite[Cheng, Ho \&
  Ruderman 1986]{1986ApJ...300..500C}), TPC/slot gap model (\cite[Dyks
  \& Rudak 2003]{2003ApJ...598.1201D}). The distinguishing features of
these pulsar models are different acceleration electric field regions
for primary particles and relevant emission mechanisms to radiate high
energy photons (\cite[Du et al.\ 2011, 2012]{vela, crab}). One of the
key discrepancies of these emission models is the one-pole or two-pole
emission pattern with two important geometry parameters: the magnetic
inclination angle $\alpha$ and the view angle $\zeta$.
%

The annular gap model was originally proposed by \cite[Qiao et
  al.\ (2004, 2007)]{qiao04,qiao07}.
The critical field lines divide the polar-cap region of a pulsar
magnetosphere into two distinct parts: the core gap region and the
annular gap region (\cite[Du et al.\ 2011, 2012]{vela, crab}). The
former gap is located around the magnetic axis and within the critical
field lines; the latter is located between the critical field lines
and the last open field lines. The width of the annular gap region is
anti-correlated with the pulsar period, it is therefore larger for
pulsars with smaller spin periods (\cite[Du et al.\ 2010]{AG10}).
The region for high energy emission in the annular gap model is
concentrated in the vicinity of the null charge surface, i.e., an
intermediate emission height, different from the outer gap model.  The
annular gap has a sufficient thickness of trans-field lines and a wide
altitude range for particle acceleration. This model combines the
advantages of the polar gap, the slot gap and the outer gap models,
and works well for pulsars with short spin periods. It is a promising
model to explain high energy emission from young and millisecond
pulsars.

\section{Modeled Results of Pulse Profiles and Spectra}

A convincing model should have a simple clear emission geometric picture
with reasonable input parameters, which can not only reproduce
multi-wavelength light curves for young pulsars but also for MSPs. 
Here we will briefly introduce our modeled results of pulse profiles
and phase-averaged spectrum for the Vela pulsar, Crab pulsar and some
millisecond pulsars. The detailed calculations in the annular gap
model can be found in \cite[Du et al.\ (2010, 2011, 2012)]{AG10,vela,crab}.

We can solve the problems of the third peak (P3) in the $\gamma$-ray
pulse profiles and the emission mechanism of GeV band for the Vela
pulsar (\cite[Du et al.\ 2011]{vela}; as shown in bottom panel of
Figure~\ref{PAS}). The GeV band emission from the Vela pulsar is
originated mainly from Synchro-curvature radiation (\cite[Zhang \&
  Cheng 1995]{syn-cur}) from primary particles, while synchrotron
radiation from secondary particles have some contributions to the
low-energy $\gamma$-ray band (e.g., $0.1 - 0.3$~GeV). Meanwhile, the
total spectrum (thick black solid line in top panel of
Figure~\ref{PAS}) is calculated in the annular gap model. It is found
that the curvature radiation and synchrotron radiation from primary
particles is mainly contributed to $\gamma$-ray band (20\,MeV to
20\,GeV); synchrotron radiation from CR-induced pairs and ICS-induced
pairs dominates the X-ray band and soft $\gamma$-ray band (100\,eV to
10\,MeV). ICS from the pairs contributes to hard TeV $\gamma$-ray band
($\sim 20$\,GeV to 400\,GeV).


\section{Conclusion}

The annular gap model is a self-consistent single-pole model not only
for young pulsars (\cite[Du et al.\ 2011, 2012]{vela,crab}), but also
for MSPs (Du et al.\ \textit{submitted}). Multi-wavelength pulsed
emission can be well explained by this model.


\end{document}